# Enhanced wakefield generation in homogeneous plasma via two co-propagating laser pulses

Abhishek Kumar Maurya[1], Dinkar Mishra[1], Bhupesh Kumar[1] Ramesh C Sharma[2], Lal C Mangal[2], Binoy K Das[2], Vijay K Saraswat[3] and Brijesh Kumar[1,4]*

[1]Department of Physics, University of Lucknow, Lucknow 226007, India

[2] Defence Research and Development Organization, Ministry of Defence, DRDO BHAWAN, New Delhi-110011

[3]Science and Technology, NITI AAYOG, PMO Office, New Delhi-110011

[4]Department of Physics & Astrophysics, University of Delhi, Delhi-110007

**Corresponding Email:** kumar0111@gmail.com

## ABSTRACT

This investigation deals with enhanced plasma wakefield amplitude generated using two co-propagating laser pulses in homogeneous plasma. The configuration consists of a seed pulse followed by a trailing pulse, both linearly polarized and sharing identical laser parameters. The enhancement in wakefield amplitude corresponding to fixed spatial separation is optimized for various pulse widths and intensities of the seed and trailing lasers. Analytical modelling and particle-in-cell simulations reveal that the maximum amplification in wakefield amplitude is obtained when spatial separation equals the plasma wavelength ($\lambda_p$). The spatial intervals between laser pulses critically influence the wakefield amplification. These findings confirm that the two co-propagating lasers scheme provides a promising route toward stronger plasma wakefield excitation, potentially important for various applications.

## I. INTRODUCTION

In plasma-based accelerators, extremely large accelerating fields can be produced without electrical breakdown- a key advantage over conventional Radio frequency (RF) accelerators. Since plasma is pre ionized, it can sustain electric field gradients up to few hundreds of GV/m which enabling the development of compact and powerful accelerators [1–4]. Among these plasma-based accelerators, the Beat Wave Accelerator (BWA) which employs two laser beams of slightly different frequencies to produce a beat frequency that resonantly drives plasma waves to efficiently accelerate particles [5–7]. When these two detuned laser beams interfere in plasma $(\Delta\omega = \omega_1 - \omega_2 = \omega_p)$, then large-amplitude plasma oscillations are produced, which lead to effective electron acceleration [5, 8,10]. In contrast, Laser Wakefield Acceleration (LWFA) uses a single intense laser pulse to excite plasma oscillations via the ponderomotive force, expelling electrons while ions remain stationary due to inertia [1, 3, 11–16]. The resulting charge separation builds strong electric fields that accelerate trapped electrons to relativistic energies [4, 6, 7, 17]. For efficient wake excitation, the laser pulse duration should satisfy the condition of $\tau_L \lesssim \lambda_p/c$ [8, 18-21]. In LWFA, an ultra-short laser pulse creates a positively charged ion channel that generates a strong longitudinal wakefield. The normalized vector potential a₀ determines the regime: linear $(a_0 \ll 1)$, nonlinear $(a_0 \approx 1)$, or relativistic $(a_0 \gg 1)$ [22-24]. The Plasma Wakefield Accelerator (PWFA), operates on a similar principle but replaces the laser pulse with a high-energy particle beam. Here, a bunch of relativistic charged particle drives plasma oscillations, while a trailing bunch gains energy by surfing the wake, achieving accelerating gradients far exceeding those of RF technology [2, 18, 19, 25-29].

In Self-Modulated Laser Wakefield Acceleration (SM-LWFA), a long laser pulse of duration $\tau_L \gg \lambda_p/c$ induces instabilities such as self-modulation and Raman scattering, consequently breaking into pulse trains that resonantly drive strong wakefields [20, 30–33]. Although SM-LWFA generates intense wakefields, the resulting beams are less controlled [33, 34]. Using two co-propagating pulses, the wakefield can be enhanced through constructive interference, which can be tuned by adjusting pulse delay and intensity [29, 35, 36]. Alternatively, counterpropagating lasers with slightly detuned frequencies can drive a beat wave at the plasma frequency, acting as a Raman seeding mechanism for controlled wake excitation [8, 37, 38]. In a homogeneous plasma, a driver (laser pulse or particle bunch) perturbs electrons, causing oscillations at the plasma frequency $\omega_p = \sqrt{(n_e e^2/m_e \epsilon_0)}$ [1, 2, 16, 39]. The corresponding wavelength of the wakefield is $\lambda_p = 2\pi c/\omega_p$, and electrons must be injected into the accelerating phase for efficient trapping [23, 36]. The dephasing length $\left(L_d \approx \lambda_p^3/\lambda_0^2\right)$ limits the maximum achievable energy, motivating the use of density tailoring and multi-stage acceleration techniques [24, 30, 38].

This work presents an analytical investigation of a recently introduced scheme for wakefield amplification, which uses two co-propagating pulses of laser that are separated by a fixed time delay distance, and is validated by Quasi-3D particle-in-cell (PIC) simulations with the Fourier-Bessel particle-in-cell (FBPIC) code [40]. This study involves a laser pulse of Gaussian profile propagating through a homogeneous plasma. The lasers are Gaussian with identical parameters in terms of polarization, frequency, and intensity, in a cold homogeneous, underdense plasma ($n_e \ll n_c$). Assuming a linear regime ($a_0 \ll 1$), with $\lambda_0 = 0.8\ \mu m$, the pulse length of seed pulsed laser such that $L \leq \lambda_p$ is generat large-amplitude wakefields. Maximum perturbation occurs when the trailing pulse length is $L \leq \lambda_p/2$, with optimal resonance $\tau_{resonant} = \pi/\omega_p$ [1, 10, 32]. The pulses propagate about one Rayleigh length without significant distortion, where $Z_R = \pi r_0^2/\lambda_0$ favors extended acceleration [22, 23, 39, 41]. Thus, efficient electron acceleration could be achieved in the weakly relativistic, linear regime using the two laser pulse scheme [24, 35, 36]. This article is organized as follows: Mathematical formulation of the problem is introduced in Section II, and further, we examine the longitudinal and transverse wakefields excited behind the seed laser pulse. Section III discusses the mechanism of laser wakefield amplification and analyzes the generated longitudinal and transverse wakefields, and their dependence on the trailing pulse and the second Gaussian laser pulse. Section IV discusses the simulation results, and we compare them with the analytical findings. Finally, Section V summarizes the main conclusion of the work.

**II. SEED PULSE DYNAMICS**

Consider a seed laser pulse of a linearly polarized electromagnetic wave which propagates along the z-axis through a uniform, pre-ionized plasma. The seed laser pulse produces an electric field given by

$$\boldsymbol{E_s}(r,z,t) = ê_x E_{0s}(r,z,t)\cos(k_0 z - \omega_0 t) \qquad (1)$$

where $E_{0s}(r,z,t)$ is represented by a slowly varying envelope of the seed pulse, $k_0$ and $\omega_0$ corresponding to the wavenumber and frequency, $ê_x$ represents its polarisation in the $x$-direction.

wakefield excitation, caused by the plasma response to a seed laser pulse in the linear regime, is described by the fluid equations (governed by the Lorentz force, continuity, and Poisson equations). Starting from the Lorentz force equation.

$$\frac{\partial v}{\partial t} = \frac{-eE_{tot}}{m} - \frac{1}{2}\nabla(v^2) + v \times \left(\nabla \times v - \frac{eB_{tot}}{mc}\right) \qquad (2)$$

where $v$ denotes the electron's fluid velocity of plasma driven by the laser perturbation field, $m$ represents the electron rest mass. Here $E_w$ and $B_w$ denotes the generated (slow oscillating) electric and magnetic wakefields. Where $E_{tot} = E + E_w$ and $B_{tot} = B + B_w$ represent the sum of the quasistatic field and the rapidly oscillating fields. As the pulse propagates, the ponderomotive force drives electrons away from regions of high laser intensity, thereby creating charge separation between the displaced electrons and the stationary ions. Such a charge imbalance gives rise to a longitudinal electrostatic wakefield $E_w$ which propagates in the wake of the laser pulse. The electron density continuity equation $n$ is

$$\frac{\partial n}{\partial t} + \nabla \cdot (nv) = 0 \qquad (3)$$

in the case of a linearly polarized laser pulse, Eq. (1), substituting into Eq. (2), the transverse electron quiver velocity is

$$v_x = ca(r,\xi,\tau)\sin(k_0\xi) \qquad (4)$$

where $a(r,\xi,\tau) = eE_{s0}(r,\xi,\tau)/mc\omega_0$ represents the normalized vector potential of the seed laser pulse, and the co-moving coordinate is $\xi = z - v_g t$, with $\tau(\approx t)$, while ignoring the direct dependence of the fluid variables on $\tau$, where $v_g$ is the group velocity of the laser pulse. This transformation is performed using the quasistatic approximation (QSA). The slowly varying axial velocity component $v_z$ evolves under the combined influence of the axial wakefield and the longitudinal ponderomotive force. The corresponding equation of motion is

$$v_g \frac{\partial v_z}{\partial \xi} = \frac{e}{m} E_{wz} + \frac{c^2}{4} \frac{\partial a^2}{\partial \xi} \qquad (5)$$

If the electron fluid is initially vortex-free, it remains vortex-free throughout the interaction $[v \times (\nabla \times v - eB_{tot}/mc) = 0]$. Where $v_z$ denotes the longitudinal component of the second-order velocity induced by the electric field. Substituting Eq. (5) into Maxwell's equations, the longitudinal wakefield equation becomes

$$\frac{v_g}{c}\frac{\partial E_{wz}}{\partial \xi} = \frac{4\pi}{c}J_z - \frac{1}{r}\frac{\partial}{\partial r}(rB_{w\theta}) \tag{6}$$

where $J_z = -env_z$ denotes the axial current density, for a uniform plasma and negligible variation of $B_{w\theta}$, Eq. (6) simplifies to

$$\left[\frac{\partial^2 E_{wz}}{\partial \xi^2} + k_p^2 E_{wz}\right] = -\frac{mc^2 k_p^2}{4e}\frac{\partial a^2}{\partial \xi} \tag{7}$$

where $k_p(=\omega_p/v_p)$ represents the plasma wavenumber and $v_p(\approx c)$, the phase velocity of the plasma wave at the plasma oscillation frequency $\omega_p(=\sqrt{n_e e^2/m_e \epsilon_0})$ Eq. (7) is a driven Helmholtz equation, where the right-hand side source term represents the ponderomotive force arising from the laser envelope. The normalised intensity profile of the laser pulse is assumed to be finite, with support only in the region $0 \leq \xi \leq L$. The formation of an axial wakefield in underdense, homogeneous plasma is determined by the dynamics of the driver, and for a sinusoidal laser pulse profile, it can be expressed as $a^2 = a_r^2 \sin^2(\pi\xi/L)$. The term $a_r^2 = a_0^2 \exp(-2r^2/r_0^2)$ represents a Gaussian transverse profile of the envelope. Here $a_0$ denotes the peak normalized vector potential, $r_0$ is the minimum spot size, and $L$ is the pulse length. The derivative of the sinusoidal pulse profile can be written as

$$\frac{\partial a^2}{\partial \xi} = a_r^2 \frac{\pi}{L}\sin\left(\frac{2\pi\xi}{L}\right) \tag{8}$$

Eq. (8) characterizes the spatial profile of the envelope along the propagation co-moving coordinate $\xi$ put in equation (7), then

$$\frac{\partial^2}{\partial \xi^2}E_{wz} + k_p^2 E_{wz} = -\frac{\epsilon k_p^2 \pi}{4L}\sin\left(\frac{2\pi\xi}{L}\right) \tag{9}$$

where $\epsilon = mc^2 a_r^2/e$ as an energy-associated amplitude parameter. The general solution of equation (9) is obtained using the Green's function that solves the operator equation $\mathcal{L} = \frac{\partial^2}{\partial \xi^2} + k_p^2$ [42] that satisfies

$$\mathcal{L}G(\xi - \xi') = \delta(\xi - \xi')$$

$$G(\xi - \xi') = \frac{\sin(k_p|\xi - \xi'|)}{k_p} \tag{10}$$

The particular solution for the wakefield is given by

$$E_{wz}(\xi) = \int_0^L G(\xi - \xi')S(\xi')d\xi' \quad (11)$$

where $S(\xi') = -\frac{\epsilon k_p^2 \pi}{4L}\sin\left(\frac{2\pi\xi'}{L}\right)$ for the point behind the pulse ($\xi < 0$), we have $|\xi - \xi'| = \xi - \xi'$. Substituting into equation (11) becomes

$$E_{wz}(\xi) = \frac{-\epsilon k_p^2 \pi}{4L}\frac{1}{k_p}\int_0^L \sin[k_p(\xi - \xi')]\sin\left(\frac{2\pi\xi'}{L}\right)d\xi' \quad (12)$$

Using the Green's function formalism, the final solution is given by

$$E_{wz}(\xi) = \frac{\epsilon k_p f}{8}\left[\sin\left(k_p(L - \xi)\right) + \sin(k_p\xi)\right] \quad (13)$$

Here, factor $f = \left[1 - \left(\frac{k_p^2 L^2}{4\pi^2}\right)\right]^{-1}$ represents a resonance correction. when $L \to \lambda_p$, the denominator of $f$ tends to zero, indicating resonant excitation of the plasma wave and hence single-pulse LWFA schemes [44, 45], Eq. (13) describes the axial seed wakefield. Taking the proper limit and keeping finite terms yields the peak wakefield amplitude, consistent with the established expression for linear wakefield generation, given by

$$E(\xi)_{wz,max} = -\frac{\epsilon \pi^2}{4\lambda_p}\cos(k_p\xi) \quad (14)$$

**III. WAKEFIELD AMPLIFICATION BY TRAILING PULSE**

The trailing laser pulse co-propagates with the leading seed pulse in the same direction, but is positioned at a distance $\Delta z$ behind it. The trailing pulse has an electric field that is linearly polarized along the $x$-axis, given by

$$\mathbf{E}_t(r, \xi', \tau') = \hat{e}_x E_{0t}(r, \xi', \tau')\cos k_0 \xi' \quad (15)$$

Here, $\xi' = z - \Delta z - v_g t$ is denotes the co-moving coordinate for the trailing pulse, $v_g$ is group velocity, and $\Delta z$ represents the distance by which the trailing pulse lags behind the seed pulse. The relative position of the trailing pulse is shifted by $\xi - \xi' = \Delta z$, showing that the trailing pulse follows the seed pulse with identical physical properties, including frequency, duration, intensity, and spatial profile. As the trailing pulse propagates in the plasma, it interacts both with the electrostatic wakefields produced by the leading seed pulse and with the electromagnetic fields that it generates itself and fields directly associated with the trailing pulse. The total electric fields and magnetic fields are given by $E'_{tot} = E_w + E_{tw}$ and $B'_{tot} = B_w + B_{tw}$ respectively. In this field configuration, the electron dynamics are governed by the Lorentz force and can be written as

$$\frac{\partial v_t}{\partial t} = \frac{-e}{m}[E_w + E_{tw}] - \frac{1}{2}\nabla(v_{tx}^2) \tag{16}$$

where $v_{tx}$ corresponds to the transverse quiver velocity of plasma electrons driven by the oscillating electric field of the trailing pulse. The quiver velocity of electrons induced by the trailing pulse is $v_{tx} = eE_t(r, \xi', \tau')/m\omega_0$, where $\omega_0$ is the laser angular frequency. The quiver motion oscillates at the laser frequency and gives rise to a ponderomotive force, represented by the gradient term $\nabla(v_{tx}^2)$ in Eq. (16), using the QSA, the slow axial motion of electrons is expressed as

$$v_g \frac{\partial v_{tz}}{\partial \xi'} = \frac{e}{m}(E_{wz} + E_{twz}) + \frac{1}{4}\frac{\partial(a_t^2 c^2)}{\partial \xi'} \tag{17}$$

Here, $v_{tz}$ represents the slow longitudinal electron velocity component, $E_{wz}$ denotes the seed wakefield, and $E_{twz}$ does the trailing pulse generate the wakefield, $a_t(r, \xi', \tau') = eE_{0t}(r, \xi', \tau')/mc\omega_0$ is the normalized amplitude of the trailing pulse. The derivative of Eq. (17) with respect to $\xi'$ and applying continuity and Poisson's equations, the governing equation for the trailing pulse wakefield becomes

$$\left[\frac{\partial^2}{\partial \xi'^2} + k_p^2\right]E_{twz} = k_p^2 E_{wz} - \frac{mc^2 k_p^2}{4e}\frac{\partial a_t^2}{\partial \xi'} \tag{18}$$

Here, $k_p = \sqrt{n_0 e^2/m\epsilon_0 c^2}$ is show the plasma wavenumber, $E_{twz}$ represents the trailing pulse contribution, the first term right-hand side of $k_p^2 E_{wz}$, couple the trailing pulse wake to the seed wake, the second term represents the direct driving force by the trailing pulse Eq. (18), which is a Helmholtz equation difference between the total wakefield and seed wakefield.

$$\delta E = E_{twz} - E_{wz} \tag{19}$$

Eq. (19), substituting in Eq. (18), then

$$\left[\frac{\partial^2}{\partial \xi'^2} + k_p^2\right]\delta E = \frac{-mc^2 k_p^2}{4e}\frac{\partial}{\partial \xi'}(a_t^2) \tag{20}$$

The corresponding symmetric Green's functions [43]

$$\left[\frac{\partial^2}{\partial \xi'^2} + k_p^2\right]G(\xi' - s) = \delta(\xi' - s) \tag{21}$$

$$G(\xi' - s) = \frac{\sin[k_p|\xi'-s|]}{2k_p} \tag{22}$$

The wakefield difference is therefore

$$\delta E(\xi') = \frac{-mc^2 k_p^2}{4e} \int G(\xi' - s) \frac{\partial a_t^2}{\partial s} ds \tag{23}$$

Assume the trailing pulse has an envelope of length $L$ centred at $\xi - \xi' = \Delta z$.

$$a_t^2(s) = \begin{cases} a_0^2, |s - \Delta z| < \frac{L}{2} \\ 0, \ Otherwise \end{cases} \tag{24}$$

The derivative of $a_t^2(s)$ is

$$\frac{\partial a_t^2}{\partial s} = a_0^2 [\delta\left(s - \Delta z - \frac{L}{2}\right) - \delta\left(s - \Delta z + \frac{L}{2}\right)] \tag{25}$$

Substituting Eq. (25) into Eq. (23) after simplification

$$\delta E(\xi') = -\frac{mc^2 k_p^2}{4e} a_0^2 \left[G\left(\xi' - \Delta z - \frac{L}{2}\right) - G\left(\xi' - \Delta z + \frac{L}{2}\right)\right]$$

$$\delta E(\xi') = -\frac{mc^2 k_p^2}{4e} \frac{a_0^2}{2k_p} [\sin(k_p\left[\xi' - \Delta z - \frac{L}{2}\right] - \sin(k_p\left[\xi' - \Delta z + \frac{L}{2}\right])]$$

$$\delta E(\xi') = -\frac{mc^2 k_p}{8e} a_0^2 [\sin(k_p\left[\xi' - \Delta z - \frac{L}{2}\right] - \sin(k_p\left[\xi' - \Delta z + \frac{L}{2}\right])] \tag{26}$$

$$\delta E(\xi') = \frac{-mc^2 k_p}{8e} a_0^2 \left(2\cos[k_p(\xi' - \Delta z)] \sin\left(\frac{k_p L}{2}\right)\right) \tag{27}$$

where $\epsilon = mc^2 a_r^2/e$, behind the pulse ($\xi' < 0$). If the trailing pulse using $\sin(k_p L/2) \approx k_p L/2$. The trailing pulse is taken sinusoidal pulse profile, represented as $a_t^2 = a_r^2 \sin^2 \pi\xi'/L$, Eq. (21) becomes

$$\delta E(\xi') = \frac{-\epsilon k_p}{4} \sin\left(\frac{k_p L}{2}\right) \cos[k_p(\xi' - \Delta z)] \tag{28}$$

Putting in Eq. (28) simplifies the expression given as

$$\delta E(\xi') = \frac{-\epsilon k_p}{4} \left(\frac{k_p L}{2}\right) \cos[k_p(\xi' - \Delta z)]$$

$$\delta E(\xi') = \frac{-\epsilon k_p^2 L}{8} \cos[k_p(\xi' - \Delta z)] \tag{29}$$

Putting $k_p = 2\pi/\lambda_p$ is the spatial frequency of plasma oscillations.

$$\delta E(\xi') = \frac{-\epsilon \pi^2 L}{2\lambda_p^2} \cos[k_p(\xi' - \Delta z)] \tag{30}$$

This is the resonant excitation maximum wake amplitude, put $L = \lambda_p/2$ (half the plasma wavelength).

$$\delta E(\xi') = \frac{-\epsilon \pi^2}{2\lambda_p^2} \frac{\lambda_p}{2} \cos[k_p(\xi' - \Delta z)]$$

$$\delta E(\xi') = \frac{-\epsilon \pi^2}{4\lambda_p} \cos[k_p(\xi' - \Delta z)] \tag{31}$$

The total wakefield consists of contributions from the seed wakefield and the trailing pulse. Peak amplitude wake of the seed pulse and $\delta E$ is the addition induced wakefield by the trailing pulse of Eqs. (14) and (31), putting in Eq. (32), then the maximum wakefield formed behind the trailing pulse is

$$E_{twz} = E_{wz} + \delta E \tag{32}$$

$$E(\xi')_{twz,max} = \frac{-\epsilon \pi^2}{4\lambda_p} \cos[k_p(\xi' - \Delta z)] - \frac{\epsilon \pi^2}{4\lambda_p} \cos[(k_p \xi)]$$

$$E(\xi')_{twz,max} = -\frac{\epsilon \pi^2}{4\lambda_p} [\cos[k_p(\xi' - \Delta z)] + \cos(k_p \xi)] \tag{33}$$

The resultant wakefield generated by the trailing pulse, and the separation between the two pulses, is $\Delta z$. The secondary trailing laser pulse modifies and enhances the plasma wakefield through its direct driving term and its coupling to the seed pulse wake, forming plasma wave amplification.

**IV. RESULTS AND DISCUSSION**

The maximum transverse wakefield amplitude, which is the longitudinal electric field $E_z$ evaluated using the analytical expression, and Quasi-3D Particle-in-cell (PIC) simulations were performed using the FBPIC code, which employs a cylindrical geometry and a moving-window configuration. Two co-propagating Gaussian laser pulses, a seed pulse (first Gaussian) and a trailing pulse (second Gaussian), were initialized in a homogeneous plasma to investigate the generation and evolution of perturbed electromagnetic fields relevant to laser–plasma amplification.

The computational domain extended 90 µm along the propagation axis (*z*) and 25 µm in the radial direction (*r*), discretized into 1200 and 100 grid cells, respectively. The corresponding spatial resolutions were Δz = 0.075 µm and Δr = 0.25 µm. Each cell contained two macroparticles per dimension in both the *z* and *r* directions, and four macroparticles per cell in the azimuthal (θ) direction. The moving window spanned from z = −10 µm to 80 µm, while the radial extent was fixed at r = 25 µm.

Open boundary conditions were applied along both *r* and *z* to suppress artificial reflections of plasma electrons at the simulation boundaries. The Gaussian driver pulses were characterized by pulse durations ranging from 10 fs to 25 fs, interacting with a uniform plasma corresponding to a plasma wavelength of approximately $\lambda_p \approx 15$ µm.

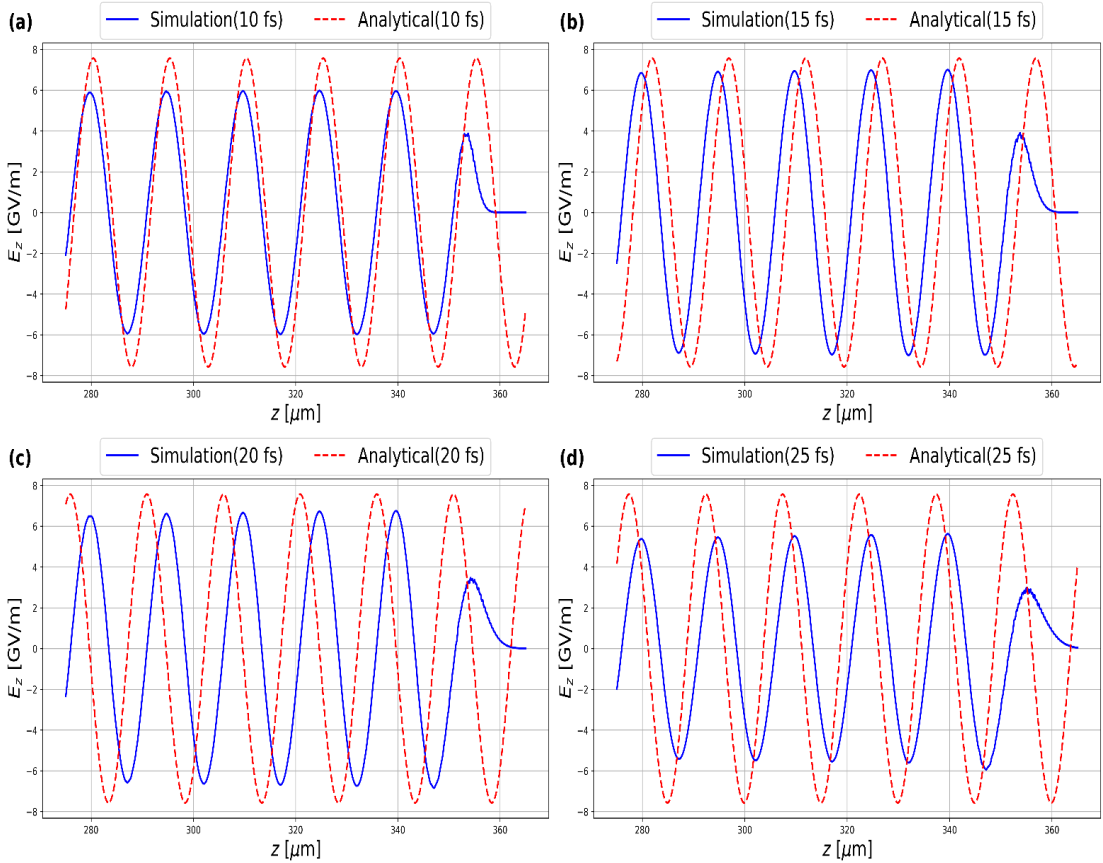

**FIG. 1.** Comparison between analytical and simulation results showing two-dimensional line plots of the longitudinal wakefield generated by the propagating in the wake of the seed pulse at time t = $9.51 \times 10^{-13}$ s. The seed pulse has parameters $a_0 = 0.3$, $r_0 = 20$ µm, $z_0 = 70$ µm, and plasma density $n_e = 4.958 \times 10^{24}\ m^{-3}$. varying pulse durations: (a) $\tau = 10$ fs, (b) $\tau = 15$ fs, (c) $\tau = 20$ fs, and (d) $\tau = 25$ fs.

Figure 1. shows a comparison between the analytical and Quasi-3D PIC simulation results for the longitudinal electric field $E_z$, illustrating the laser wakefield generation process driven by Gaussian pulse (Seed pulse) in an underdense plasma. The seed pulse initiates the plasma

oscillation, with seed pulse parameters that are $a_0 = 0.3$, $r_0 = 20$ μm, $z_0 = 70$ μm, and plasma density $n_e = 4.958 \times 10^{24}$ $m^{-3}$. For the best case shown in Figure 1(b), which corresponds to a pulse duration of $\tau = 15$ fs. The analytical configuration and simulation yield maximum wakefield amplitudes of 7.43 GV/m and 7.10 GV/m, respectively, demonstrating strong agreement within the linear regime. The results of wakefield generated with varying pulse durations are presented in Figs. 1(a)-1(d).

**TABLE I.** shows the comparison between analytical and simulation results for the laser wakefield generated at $a_0 = 0.3$, normalized vector potentials of the Seed pulse.

| Pulse duration ($\tau$) fs | Analytical ($GV/m$) | Simulation ($GV/m$) |
|---|---|---|
| 10 | 7.42 | 6.00 |
| 15 | 7.43 | 7.10 |
| 20 | 7.45 | 6.81 |
| 25 | 7.46 | 5.65 |

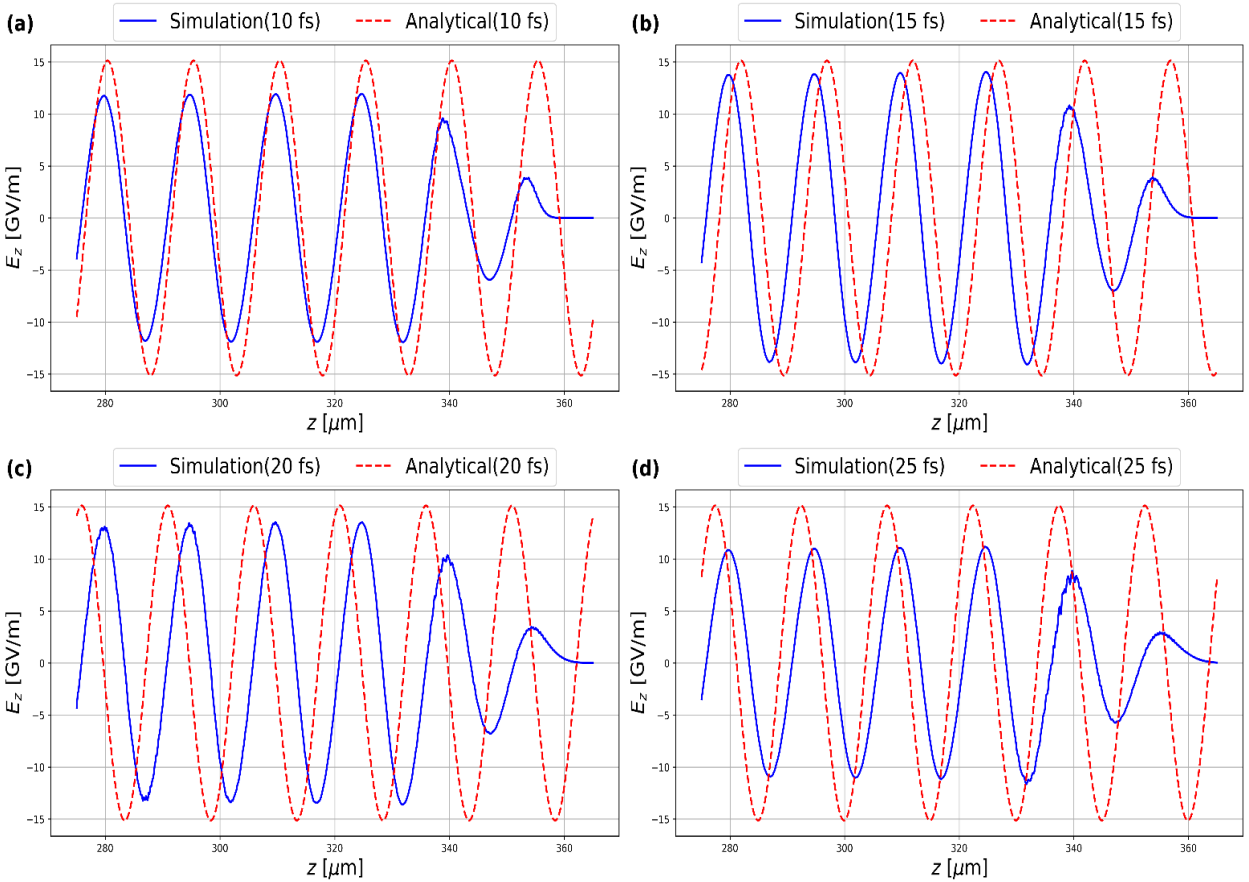

**FIG. 2.** Comparison between analytical and simulation results showing two-dimensional line plots of the enhanced longitudinal wakefield generated by the trailing pulse propagating in the wake of the seed pulse at time t $= 9.51 \times 10^{-13}$ s. The seed pulse has parameters $a_0 = 0.3$, $r_0 = 20$ μm, $z_0 = 70$ μm, and plasma density $n_e =$

$4.958 \times 10^{24}\ m^{-3}$. The trailing pulse possesses identical parameters but is centered at $z_0 = 55\ \mu m$ varying pulse durations: (a) $\tau = 10$ fs, (b) $\tau = 15$ fs, (c) $\tau = 20$ fs, and (d) $\tau = 25$ fs.

Figure 2 presents a comparison between the analytical predictions and Quasi-3D PIC simulations results for the longitudinal electric field $E_z$, illustrating the wakefield amplification process driven by two co-propagating Gaussian laser pulses in an under-dense plasma. The seed pulse initiates the plasma oscillation, while the trailing pulse reinforces the wakefield within the pre-formed plasma wave. The seed pulse parameters are $a_0 = 0.3$, $r_0 = 20\ \mu m$, $z_0 = 70\ \mu m$, and plasma density $n_e = 4.958 \times 10^{24}\ m^{-3}$. The trailing pulse, positioned at $z_0 = 55\ \mu m$, maintains identical parameters, establishing a pulse separation of $\Delta z \approx 15\ \mu m$. For the best case shown in Figure 2(b), corresponding to a pulse duration of $\tau = 15$ fs, the analytical configuration and simulation yield maximum wakefield amplitudes of 14.95 GV/m and 14.25 GV/m, respectively, demonstrating strong agreement within the linear regime.

**TABLE II.** shows the Comparison between analytical and simulation results for laser wakefield amplification at $a_0 = 0.3$, normalized vector potentials of the trailing pulse.

| Pulse duration ($\tau$) fs | Analytical Trailing pulse amplification ($GV/m$) | Simulation Trailing pulse amplification ($GV/m$) |
|---|---|---|
| 10 | 14.93 | 12.06 |
| 15 | 14.95 | 14.25 |
| 20 | 14.96 | 13.63 |
| 25 | 14.98 | 11.30 |

The results of wakefield amplification with varying pulse durations are presented in Figs. 2(a)-2(d). This trend indicates that shorter pulse durations are more effective in resonantly driving and amplifying the plasma wave due to better synchronization with the plasma oscillation period ($\lambda_p \approx 15\ \mu m$), the small discrepancy between analytical and simulation amplitudes can be attributed to multidimensional effects, such as transverse diffraction and finite spot-size corrections, which are not fully captured by the one-dimensional analytical formulation. Overall, the results confirm that efficient wakefield amplification occurs when the trailing pulse is temporally synchronized with the plasma wave excited by the seed laser pulse, leading to constructive interference, which subsequently enhanced longitudinal field strength.

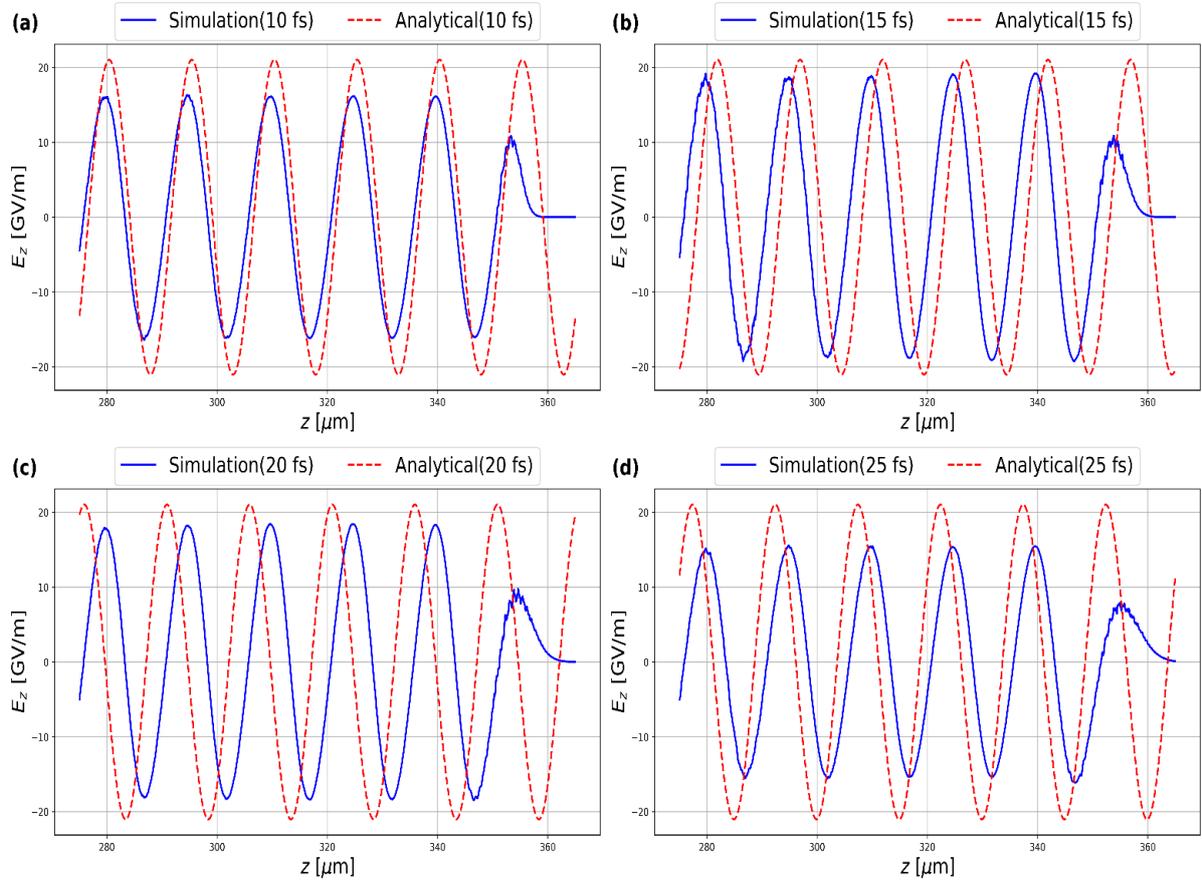

FIG. 3. Comparison between analytical and simulation results showing two-dimensional line plots of the enhanced longitudinal wakefield generated by the propagating in the wake of the seed pulse at time t = $9.51 \times 10^{-13}$s. The seed pulse has parameters $a_0 = 0.5$, $r_0 = 20$ μm, $z_0 = 70$ μm, and plasma density $n_e = 4.958 \times 10^{24}\ m^{-3}$. varying pulse durations: (a) $\tau = 10$ fs, (b) $\tau = 15$ fs, (c) $\tau = 20$ fs, and (d) $\tau = 25$ fs.

Figure 3. presents a comparison between the analytical and Quasi-3D PIC simulations results for the $E_z$, a longitudinal electric field, illustrating the wakefield enhanced process driven by Gaussian laser pulses (Seed pulse) in an underdense plasma. The seed pulse initiates the plasma oscillation, with seed pulse parameters that are $a_0 = 0.5$, $r_0 = 20$ μm, $z_0 = 70$ μm, and plasma density $n_e = 4.958 \times 10^{24}\ m^{-3}$. For the best case shown in Figure 1(b), corresponding to a pulse duration of $\tau = 15$ fs, the analytical configuration and simulation yield maximum wakefield amplitudes of 20.50 GV/m and 19.30 GV/m, respectively, demonstrating strong agreement within the linear regime. The results of wakefield enhancement with varying pulse durations are presented in Figs. 3(a)-3(d)

**TABLE III.** shows the comparison between analytical and simulation results for laser wakefield generated at $a_0 = 0.5$, normalized vector potentials of the Seed pulse.

| Pulse duration ($\tau$) fs | Analytical ($GV/m$) | Simulation ($GV/m$) |
|---|---|---|
| 10 | 20.45 | 16.38 |
| 15 | 20.50 | 19.30 |
| 20 | 20.55 | 18.53 |
| 25 | 20.60 | 15.62 |

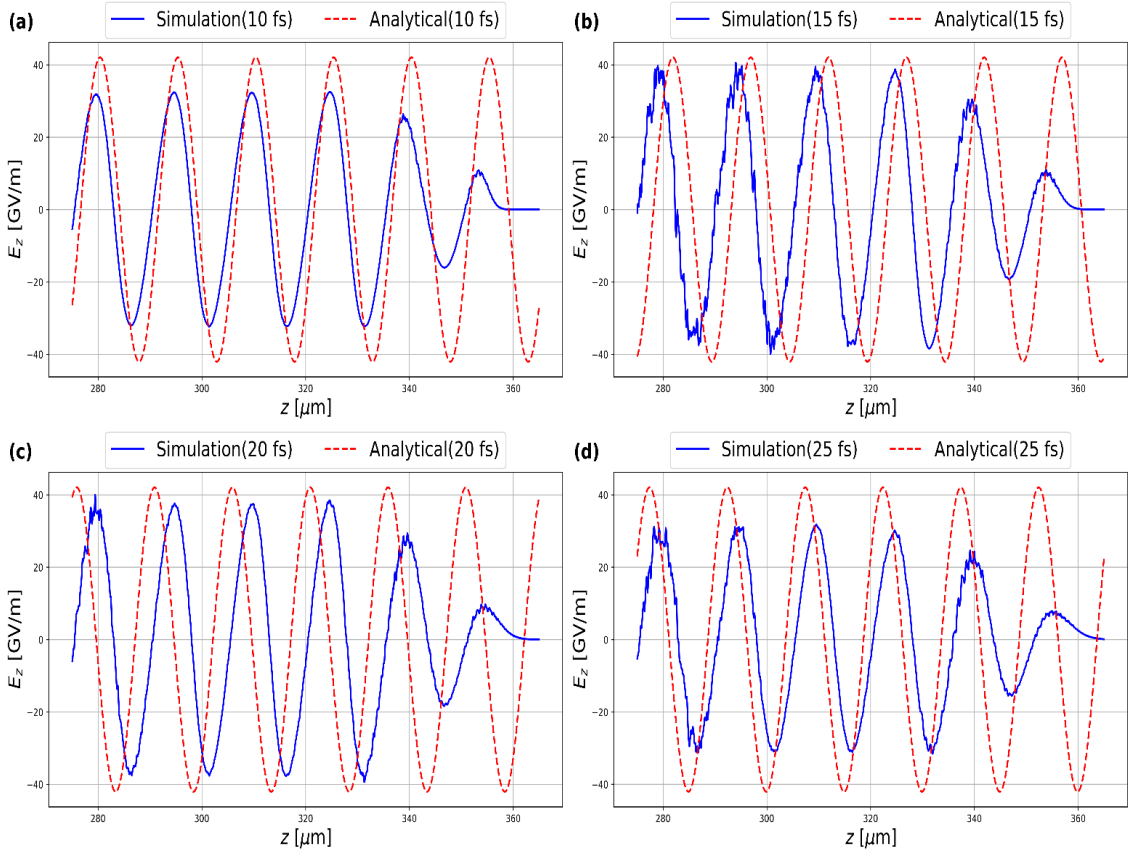

**FIG. 4.** Comparison between analytical and simulation results showing two-dimensional line plots of the enhanced longitudinal wakefield generated by the trailing pulse propagating in the wake of the seed pulse at time t $= 9.51 \times 10^{-13}$s. The seed pulse has parameters $a_0 = 0.5$, $r_0 = 20$ μ$m$, $z_0 = 70$ μ$m$, and plasma density $n_e = 4.958 \times 10^{24}\ m^{-3}$. The trailing pulse possesses identical parameters but is centred at $z_0 = 55$ μ$m$ varying pulse durations: (a) $\tau = 10$ fs, (b) $\tau = 15$ fs, (c) $\tau = 20$ fs, and (d) $\tau = 25$ fs.

Figure 4 shows a detailed comparison between analytical predictions and Quasi-3D PIC simulations results for the $E_z$ longitudinal electric field in the case of a higher normalized vector potential, $a_0 = 0.5$. Both the analytical and simulation results demonstrate a significant amplification of the plasma wakefield when the trailing Gaussian pulse follows the seed pulse with a spatial separation of $\Delta z = 15\ \mu$m. For the best configuration shown in Fig. 4(b), corresponding to a pulse duration of $\tau = 15$ fs, the maximum wakefield amplitudes are 41.52 GV/m (analytical) and 39.70 GV/m (simulation). The increase in $a_0$ compared with the previous case $(a_0 = 0.3)$ results in a stronger ponderomotive force, thereby generating a higher-amplitude plasma density perturbation and, consequently, a stronger longitudinal field.

**TABLE IV.** shows the comparison between analytical and simulation results for laser wakefield amplification at $a_0 = 0.5$, normalized vector potentials of the trailing pulse.

| Pulse duration ($\tau$) fs | Analytical Trailing pulse amplification ($GV/m$) | Simulation Trailing pulse amplification ($GV/m$) |
|---|---|---|
| 10 | 41.46 | 32.80 |
| 15 | 41.52 | 39.70 |
| 20 | 41.57 | 38.60 |
| 25 | 41.60 | 32.00 |

For $a_0 = 0.3$ and $a_0 = 0.5$, across all cases, both the analytical calculations and the simulation results consistently show that the introduction of a trailing laser pulse nearly doubles the wakefield amplitudes generated by the seed pulse. This amplification is approximately 100 percent, indicating a strong enhancement of the longitudinal electric field due to constructive interference between the seed and trailing pulses.

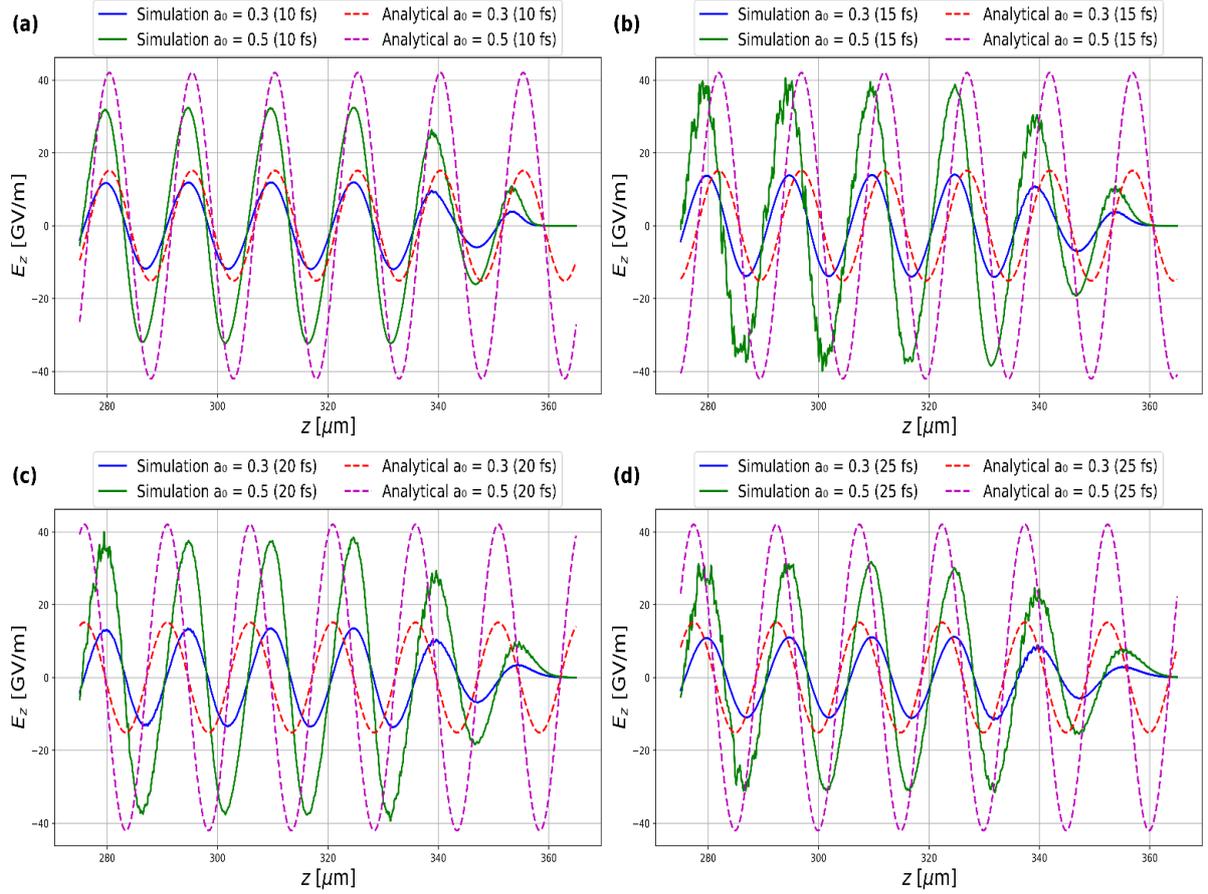

**FIG. 5.** Comparison between analytical and simulation two-dimensional line plots of the longitudinal electric field $E_z$ illustrating laser wakefield amplification at $t = 9.51 \times 10^{-13}\,s$ (corresponding to iteration 3800). Both the seed and trailing Gaussian laser pulses have identical parameters and are separated by $\Delta z \approx 15$ μm. The comparison of analytical and simulation results is shown for different normalized vector potentials of seed and the trailing pulse, $a_0 = 0.3$ and $a_0 = 0.5$. The Gaussian pulse durations are varied as follows: (a) $\tau = 10$ fs, (b) $\tau = 15$ fs, (c) $\tau = 20$ fs, and (d) $\tau = 25$ fs.

The results of wakefield amplification at $a_0 = 0.5$, with varying pulse durations, are presented in Figs. 4(a)-4(d). These results show a gradual reduction in the simulated wakefield amplitude with increasing pulse duration, indicating that shorter pulses are more effective in resonantly driving the plasma wave. This behaviour arises because a shorter pulse duration better matches the plasma period $T_p = 2\pi/\omega_p$, maintaining constructive interference between the fields generated by the seed and trailing pulses.

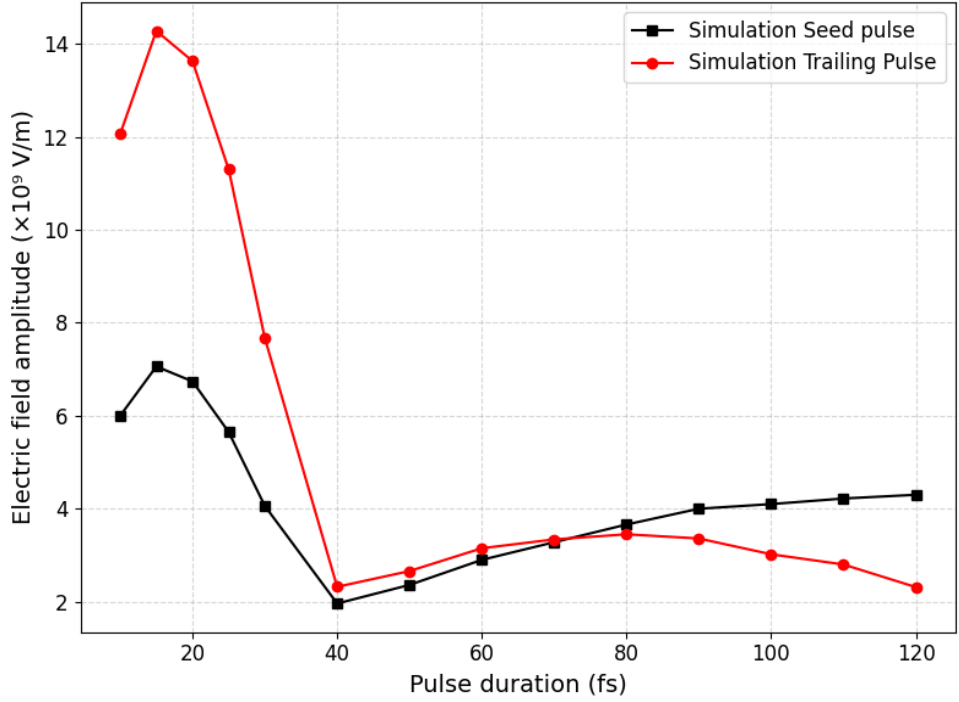

**FIG. 6.** Simulation of two-dimensional line plots of the longitudinal electric field $E_z$ illustrating laser wakefield amplification at $t = 9.51 \times 10^{-13}$s (corresponding to iteration 3800). The Gaussian laser pulse duration is varied from 10 fs to 120 fs. The seed pulse has parameters $a_0 = 0.3$, $r_0 = 20$ μm, $z_0 = 70$ μm, and $n_e = 4.958 \times 10^{24}$m$^{-3}$, while the trailing pulse has identical parameters and is centered at $z_0 = 40$ μm.

The seed and trailing laser pulses are configured such that in-phase with the spatial separation between the two pulses is approximately equal to the plasma wavelength, $\lambda_p \approx 15$ μm (Fig. 5). Consequently, the wakefield excited by the trailing pulse adds coherently to that generated by the seed pulse, resulting in constructive interference and a significant amplification of the total plasma wakefield amplitude. This coherent superposition enhances plasma oscillations and promotes efficient energy transfer from the laser field to the plasma wave. When the pulse separation satisfies $L \leq \lambda_p$, the wakefields from both pulses reinforce each other, leading to optimal laser wakefield amplification. Variation in pulse duration of the seed and trailing pulses, while maintaining identical parameters and a separation $\Delta z \approx \lambda_p$, does not degrade the constructive interference condition, thereby preserving efficient wakefield amplification.

Figure 6 illustrates the phase relationship between two co-propagating laser pulses separated by $\Delta z \approx 2\lambda_p$, where $\lambda_p$ is the plasma wavelength. The plasma oscillation period $T_p$ represents the time required for electrons to complete one full oscillation in the restoring electric field of the plasma and is defined as $T_p = \frac{2\pi}{\omega_p}$, where $\omega_p = \sqrt{\frac{n_e e^2}{\epsilon_0 m_e}}$. For the plasma parameters used in this study, $T_p \approx 50$ fs. The phase of the plasma oscillation depends primarily on the plasma frequency $\omega_p$ rather than on the pulse duration $\tau$, which ensures that the trailing pulse consistently arrives at the same oscillation phase relative to the seed pulse.

Under the present simulation conditions, the laser pulse length is chosen such that $L \approx \lambda_p/2$, corresponding to half the plasma wavelength. This configuration satisfies the resonance criterion for maximum wake excitation, where $\tau \approx T_p/2$. Hence, $\tau \approx \frac{1}{2}\left(\frac{2\pi}{\omega_p}\right) \leq \frac{\pi}{\omega_p} \approx 25$ fs. This condition represents the optimal, or resonant, regime in which the wakefield amplitude is maximized through constructive interference between the seed and trailing pulses. In this range (10 fs $\leq \tau \leq$ 25 fs), the plasma oscillations are coherent, resulting in strong amplification of the wakefield.

When the pulse duration increases beyond $\tau > 50$ fs, phase mismatch occurs between the laser envelope and the plasma oscillation, causing the wakefield to move out of phase with the driving field. In such cases, the efficiency of energy transfer decreases, and the wake amplitude diminishes. Moreover, when the pulse length satisfies $L \approx (2n + 1)\frac{\lambda_p}{2}$, the wakefield contributions from successive oscillations interfere destructively [1,46]. This destructive interference is observed, for instance, in the case of an 80 fs pulse (corresponding to a spatial extent of $\approx 24$ μm), where the wakefield amplitude is significantly reduced. Thus, the phase-matching condition between the laser pulse and plasma oscillation is crucial for achieving optimal wakefield amplification and maintaining coherent plasma wave excitation.

## V. SUMMARY AND CONCLUSION

This study presents an analytical and simulation investigation of laser wakefield amplification driven by two co-propagating Gaussian laser pulses in a homogeneous plasma. Using the quasistatic approximation, the driven Helmholtz equation for the longitudinal wakefield was derived and solved via the Green's function method, which shows that the wake amplitude depends strongly on the pulse separation and normalized vector potential $a_0$. Analytical and Quasi-3D PIC simulations reveal that when the pulse separation is approximately equal to the plasma wavelength ($\Delta z \approx \lambda_p$), the wakefields generated by the seed and trailing pulses add coherently, leading to constructive interference and enhanced plasma oscillations. The maximum amplification occurs under the resonant condition $\tau \approx T_p/2 = \pi/\omega_p \approx 25$ fs, corresponding to half the plasma oscillation period. For longer pulses ($\tau > 50$ fs), phase mismatch leads to reduced wake amplitudes and eventual destructive interference.

Simulation results validated the analytical trends, and showing maximum longitudinal field amplitudes of 14.95 GV/m and 14.25 $GV/m$ for $a_0 = 0.3$, and 41.52 GV/m and 39.70 $GV/m$ for $a_0 = 0.5$, consistent with the expected $a_0^2$ scaling. The enhanced longitudinal field if effectively trapped an injected low energy electron behind the trailing pulse and then it can accelerate to higher energies than in the single-pulse case. In conclusion, two co-propagating Gaussian laser schemes provide an efficient and controllable route to amplify plasma wakefields through coherent superposition and phase matching. The results establish that maintaining $\Delta z \approx \lambda_p$ and $\tau \approx T_p/2$ enables optimal wake excitation and energy transfer, offering a promising framework for next-generation laser–plasma accelerators.

**DATA AVAILABILITY STATEMENT**

The data that support the findings of this study are available within the manuscript itself. All numerical results, analytical derivations, and simulation parameters required to reproduce the reported analyses are presented in the corresponding sections and figures of this paper.